\documentclass[12pt,a4paper]{article}
\newcommand\be{\begin{equation}}
\newcommand\ee{\end{equation}}
\newcommand\bea{\begin{eqnarray}}
\newcommand\eea{\end{eqnarray}}
\newcommand\ket[1]{|#1\rangle}
\newcommand\bra[1]{\langle #1|}
\newcommand\bk[2]{\langle #1|#2\rangle}
\newcommand\pty{{$\cal PT$}-symmetry}
\newcommand\ptc{{$\cal PT$}-symmetric}
\newcommand\pti{{$\cal PT$}-invariant}
\newcommand\pte{{$\cal PT$}-invariance}
\title{Completeness and Orthonormality in \ptc\ Quantum Systems}
\author{Stefan Weigert \\
HuMP (Hull Mathematical Physics)\\
              Department of Mathematics, University of Hull\\
              Cottingham Road, UK-Hull HU6 7RX,
              United Kingdom\\
\tt s.weigert@hull.ac.uk}
\date{June 2003}
\begin{document}
\maketitle
\begin{abstract}
Some \ptc\ non-hermitean Hamiltonians have only real eigenvalues.
There is numerical evidence that the associated \pti\ energy
eigenstates satisfy an unconventional completeness relation. An
{\em ad hoc} scalar product among the states is positive definite
only if a recently introduced `charge operator' is included in its
definition. A simple derivation of the conjectured completeness
and orthonormality relations is given. It exploits the fact that
\pty\ provides an additional link between the eigenstates of the
Hamiltonian and those of its adjoint, which form a dual pair of
bases. The `charge operator' emerges naturally upon expressing the
properties of the dual bases in terms of {\em one} basis only.
\end{abstract}
%
%
Hermitean operators have real eigenvalues while non-hermitean ones
may have complex eigenvalues. Numerical and analytical results
indicated the possibility to compensate the non-hermiticity of a
Hamiltonian by the presence of an additional symmetry
\cite{bender+98}. The spectra of many  non-hermiteans Hamiltonians
$\hat H$ are indeed {\em real} \cite{spectra} if they are
invariant under the combined action of self-adjoint parity $\cal
P$ and time reversal $\cal T$,
\be
[ \hat H , {\cal P T} ] = 0 \, ,
\label{ptsymm} \ee
and if the energy eigenstates are invariant under the operator
${\cal PT}$. Pairs of {\em complex conjugate} eigenvalues are
compatible with \pty\ as well but the eigenstates of $\hat H$ are
no longer invariant under $\cal PT$. It is possible to explain
these observations by the concept of {\em pseudo-hermitean}
operators \cite{multimostafazadeh} which satisfy
\be
\eta \hat H \eta^{-1} = {\hat H}^\dagger \, ,
\label{pseudo-herm} \ee
following from Eq. (\ref{ptsymm}) with $\eta = {\cal P}$. Wigner's
representation theory of anti-linear operators \cite{wigner60}
provides an alternative explanation if applied to the operator
${\cal PT}$ \cite{weigert02}. What is more, the group theoretical
approach explains the fate of energy eigenstates if they are {\em
not} invariant under the action of ${\cal PT}$, and a complete
classification of \pti\ subspaces emerges.

\ptc\ systems possess at least two other intriguing features.
First, the eigenstates of \ptc\ non-hermitean Hamiltonians (with
real eigenvalues only) do not satisfy the standard completeness
relations. Numerical evidence \cite{bender+02} suggests that one
has instead
\be
\sum_{n} (-1)^n \phi_n(x) \phi_n (y) = \delta(x-y)\, ,
\label{positioncomp1} \ee
the functions $\phi_n(x)\equiv \bk{x}{E_n}$ being energy
eigenstates of a particle on the real line subjected to a \ptc\
potential such as $V(x) = x^2 (ix)^\nu, \nu \geq 0$
\cite{multibender}. Whether the completeness (\ref{positioncomp1})
relation is valid has been called a `major open mathematical
question for \ptc\ Hamiltonians' \cite{majorproblem}. Second, a
`natural inner product' of functions $f(x)$ and $g(x)$ associated
with \ptc\ systems has been proposed \cite{bagchi+01},
\be
(f,g) = \int dx [{\cal PT} f(x)] g(x) \, ,
\label{inner1} \ee
where the integration is along an appropriate path, possibly in
the complex-$x$ plane \cite{bender+02}. This scalar product
implies that energy eigenstates can have a negative norm,
\be
(\phi_m,\phi_n) = (-1)^n \delta_{mn} \, .
\label{negnorm} \ee
which makes it difficult to maintain the familiar probabilistic
interpretation of quantum theory \cite{bagchi+01} and gave rise to
discussions about the state space of \ptc\ systems
\cite{japaridze02}.

In an attempt to base an extension of quantum mechanics
\cite{bender+02} on systems with \pty\, a remedy against the
indefinite metric in Hilbert space has been proposed in the form
of a linear `charge operator' ${\cal C}$. Its  position
representation is given by
\be
{\cal C} (x,y) = \sum_n \phi_n(x) \phi_n (y) \, .
\label{chargepos} \ee
Then, the redefined inner product
\be
\langle f | g\rangle = \int_C dx [{\cal CPT} f(x)] g(x) \, ,
\label{inner2} \ee
is positive definite, and the completeness relation
(\ref{positioncomp1}) turns into
\be
\sum_n [ {\cal CPT} \phi_n(x)] \phi_n (y) = \delta(x-y)\, .
\label{positioncomp2} \ee
These relations are also consistent with results obtained for
pseudo-hermitean operators
\cite{multimostafazadeh,mostafazadeh03}.

The purpose of this contribution is, first, to prove that
relations such as (\ref{positioncomp1}) exist for all \ptc\ system
with real eigenvalues. Second, the origin of the operator ${\cal
C}$ will be identified, which directly explains both why Eq.
(\ref{inner2}) defines indeed a positive inner product and why Eq.
(\ref{positioncomp2}) is a valid completeness relation. To cut a
long story short, the last two equations (as well as
(\ref{positioncomp1}) and (\ref{inner1})) are nothing but
bi-orthonormality and completeness for a pair of dual bases
associated with $\hat H$. It is due to the system's \pty\ and the
occurrence of real eigenvalues only that these two relations
acquire a special form which involves the elements $\{ \phi_n
(x)\}$ of {\em one} basis only.

%
%
Consider a (diagonalizable) {\em non-hermitean} Hamiltonian $\hat
H$ with a discrete spectrum \cite{wong67}. The operators $\hat H$
and and its adjoint $\hat H^\dagger$ have complete sets of
eigenstates:
\be \label{nonhermH}
\hat H \ket{E_n} = E_n \ket{E_n} \, , \quad
 \hat H^\dagger \ket{E^n} = E^n \ket{E^n} \, , \quad n= 1,2, \dots \, ,
\ee
with, in general, complex conjugate eigenvalues, $E^n = E_n^*$.
The eigenstates constitute {\em bi-orthonormal} bases in $\cal H$
with two resolutions of unity,
\be \label{completeness}
\sum_n \ket{E^n}\bra{E_n}
 = \sum_n \ket{E_n}\bra{E^n}
 = \hat I \, ,
\ee
and as dual bases, they satisfy orthonormality relations,
\be\label{dual}
\bra{E^n} E_m \rangle = \bra{E_m} E^n \rangle = \delta_{nm} \, ,
\quad m,n = 1,2, \ldots
\ee
{\em A priori}, nothing is known about scalar products such as
$\bra{E_n} E_m \rangle$.

Consider now a \pti\ Hamiltonian, i.e., Eq. (\ref{ptsymm}) holds,
and assume all eigenvalues to be real and non-degenerate. Multiply
the first equation of (\ref{nonhermH}) with the operator $\cal PT$
so that
\be \label{ptstates}
\hat H \left({\cal PT} \ket{E_n}\right)
          = E_n \left({\cal PT} \ket{E_n}\right) \, .
\ee
Consequently, the state ${\cal PT} \ket{E_n}$ must equal
$\ket{E_n}$ apart from a factor $d_n$. Since $({\cal PT})^2
\ket{E_n} = \ket{E_n} = |d_n|^2 \ket{E_n}$, the numbers $d_n$
equal phase factors $e^{i\varphi_n}$, say. Redefining $\ket{E_n}
\rightarrow e^{i\varphi_n/2} \ket{E_n}$ implies---as is
well-known---that one can always write
\be
{\cal PT} \ket{E_n} = \ket{E_n}  \quad \mbox{ or } \quad \phi_n^*
(-x) = \phi_n (x) \, .
\label{pteigeneq}\ee

\pty\ of a non-hermitean Hamiltonian $\hat H$ leads to particular
relation between the operator and its adjoint $\hat H^\dagger$. As
mentioned earlier, the adjoint of $\hat H$ can be obtained from
applying parity to it,
\be \label{hamsym}
{\hat H}^\dagger = {\cal P} \hat H {\cal P} \, .
\ee
It will be shown now that a simple relation between the states
$\ket{E_n}$ and $\ket{E^n}$ results, {\em viz.},
\be \label{newprop}
 \ket{E^n} = s_n {\cal P} \ket{E_n} \, , \quad
 s_n = \pm 1 \, .
\ee
This relation is crucial to derive the numerically observed
completeness and orthogonality relations. To see that
(\ref{newprop}) holds, an argument similar to the derivation of
Eq. (\ref{pteigeneq}) will be given. Write $\hat H^\dagger = {\cal
P} \hat H {\cal P}$ in the second equation of (\ref{nonhermH}),
multiply it with $\cal P$, use ${\cal P}^2=\hat I$ and recall that
$E^n = E_n^* = E_n$:
\be \label{crucialrel}
\hat H \left( {\cal P}\ket{E^n}\right)
     = E_n \left({\cal P} \ket{E^n}\right) \, .
\ee
Comparison with the first equation of (\ref{nonhermH}) shows that
the states  ${\cal P}\ket{E^n}$ and $\ket{E_n}$ are both
eigenstates of $\hat H$, with the {\em same} non-denerate
eigenvalue $E_n$. Consequently, they must be proportional to each
other,
\be \label{newpropc}
 \ket{E^n} = c_n {\cal P} \ket{E_n} \, , \quad
 c_n \in {\sf C}\, .
\ee
The numbers $c_n$ must, in fact, be {\em real} since the states $
\ket{E_n}$ and $\ket{E^n}$ are a normalized pair: using ${\cal
P}^2 = \hat I$ and (\ref{newprop}) implies
\be
1 = \bra{E^n} E_n\rangle
  = \bra{E^n} {\cal P}^2 \ket{ E_n}
  = c^*_n c^{-1}_n \bra{E_n} E^n\rangle = c^*_n c^{-1}_n \, ,
\label{reality} \ee
that is, $c_n = c_n^*$. Furthermore, the dual bases can always be
chosen in such a way that the numbers $c_n$ will take the values
$\pm 1$. To see this, multiply each side of (\ref{newpropc}) with
its own adjoint, giving $\bk{E^n}{E^n} = c^2_n \bk{E_n}{E_n}$, or
\be
c_n = s_n \left( \frac{ \bk{E^n}{E^n}}{\bk{E_n}{E_n}}
\right)^{1/2} \, , \quad s_n = \pm 1\, ,
\label{cnmod} \ee
consistent with (\ref{reality}) because the scalar products are
positive. The square root can always be given the value one by
rescaling the eigenstates of $\hat H$ and ${\hat H}^\dagger$. For
each dual pair, let
\be
\ket{E_n} \rightarrow \lambda_n \ket{E_n} \, \quad \mbox{ and } \quad
\ket{E^n} \rightarrow \lambda_n^{-1} \ket{E^n} \, , \quad
0< \lambda_n < \infty \, ,
\label{scale} \ee
a transformation which does not change orthonormality of the bases
since $\bk{E_n}{E^m}$ remains invariant. Eq. (\ref{cnmod}),
however, turns into \be
c_n = s_n \left( \frac{1}{\lambda_n^4} \frac{ \bk{E^n}{E^n}}{\bk{E_n}{E_n}}
   \right)^{1/2} \equiv s_n  \quad \mbox{ if } \quad
 \lambda_n = \left( \frac{ \bk{E^n}{E^n}}{\bk{E_n}{E_n}} \right)^{1/4} \, .
\label{cnmodscaled} \ee
The {\em signature} $s = (s_1,s_2, \ldots)$ depends on the actual
Hamiltonian as a discussion of finite-dimensional \ptc\ systems
\cite{bender+03} shows. Here is a simple way to calculate the
numbers $s_n$ once the eigenfunctions $\phi_n(x) = \bk{x}{E_n}$ of
a Hamiltonian with \pty\ have been determined. Multiply Eq.
(\ref{newprop}) with $\bra{E_n}$ and solve for $s_n \equiv
s_n^{-1}$:
\be
s_n = \bra{E_n} P \ket{E_n} \, .
\label{valuesn} \ee
%

%
%
Using (\ref{newprop}), it is straightforward to derive
completeness relations which involve the states of {\em one} basis
only. Rewrite (\ref{completeness}) by means of (\ref{newprop}) as
\be \label{newcompleteness}
 \sum_n \ket{E_n}\bra{E^n}
     = \sum_n s_n \ket{E_n}\bra{E_n} {\cal P} = \hat I \, ,
\ee
and take its matrix elements in the position representation
\be
\sum_n s_n \phi_n (x) \phi^*_n(-y)
    = \sum_n s_n \phi_n (x) \phi_n(y) =\delta (x-y) \, ,
\label{unconcompl} \ee
where $\cal PT$-invariance (\ref{pteigeneq}) has been used. The
result agrees with the expression (\ref{positioncomp1}) if $s_n =
(-1)^n$. In a similar way, one can derive a completeness relation
for the eigenstates of ${\hat H}^\dagger$,
\be
\sum_n s_n \phi^n (x) \phi^n(y) = \delta (x-y) \, .
\label{unconcompldual} \ee
%

The orthonormality condition for dual states turns into a relation
which has been interpreted as the existence of a non-positive
scalar product among the eigenstates of $\hat H$. Simply write the
scalar product (\ref{dual}) in the position representation, using
(\ref{newprop}) and \pte ,
\bea
\bk{E^n}{E_m} &=& s_n \bra{E_n} {\cal P} \ket{E_m}
                  = s_n \int dx \, \phi^*_n (-x) \phi_m (x)   \nonumber\\
              &=& s_n \int dx \, \phi_n (x) \phi_m (x)
              = \delta_{nm} \, ,
\label{scprtrf} \eea
or, using the notation from (\ref{inner1}),
\be
(\phi_n,\phi_m) = s_n \delta_{nm} \, ,
\label{inner3} \ee
which is again consistent with $s_n = (-1)^n$.
%
%

Suppose we wanted to write an operator version of (\ref{newprop}).
Define an operator ${\cal C}_s$ by
\be
{\cal C}_s = \sum_k s_k \ket{E_k} \bra{E^k} \, .
\label{Cs} \ee
Its eigenstates are $\ket{E_n}$ since
\be
{\cal C}_s \ket{E_n}
  = \sum_k s_k \ket{E_k} \bk{E^k}{E_n}
  = s_n \ket{E_n}  \, ,
\label{CSeigen} \ee
and its eigenvalues $s_n$ coincide indeed with the signs of the
`${\cal PT}$-norm,' a property of the `charge operator' $\cal C$
pointed out in \cite{bender+02}. Writing
\be \label{newpropop}
 \ket{E^n} = s_n {\cal P} \ket{E_n}
           = {\cal P} {\cal C}_s \ket{E_n} \, ,
\ee
one can transform the scalar product of dual states, using
(\ref{pteigeneq}) twice,
\bea
\bk{E_m}{E^n} &=&  \bra{E_m} {\cal P} {\cal C}_s \ket{E_m}
               =  \bra{E_m} {\cal P} \int dx \, \ket{x} \bra{x}
                                      {\cal C}_s \ket{E_m}
                                      \nonumber \\
              &=&  \int dx \, \phi_m^* (-x) \, {\cal C}_s \,
                    \phi_n (x)
               =   \int dx \, \phi_m (x) [{\cal C}_s {\cal P} {\cal T}
                          \phi_n (x)]
              = \delta_{nm} \, .
\label{scprtrfnew} \eea
Defining ${\cal C}_s = {\cal C}$ if $s_n = (-1)^n$, this equation
justifies (\ref{inner2}) for energy eigenstates. Furthermore, the
first completeness relation in (\ref{completeness}) implies
through (\ref{newpropop}) that
\bea \label{cptcompleteness}
\delta(x-y) &=& \sum_n \bra{x} {\cal P} \ket{E^n} \bra{E_n}
                                {\cal P}\ket{y} \nonumber \\
            &=& \sum_n {\cal C}_s \phi_n(x) \phi_n^* (-y)
            = \sum_n \, [{\cal C}_s {\cal PT} \phi_n(x)] \phi_n (y) \, ,
\eea
which reproduces (\ref{positioncomp2}), identical to Eq. (13) of
\cite{bender+02}. By taking matrix elements of Eq. (\ref{Cs}), the
position representation of the operator ${\cal C}_s(x,y)$ is found
to agree with (\ref{chargepos}).

In summary, it has been shown that the dual bases of \ptc\ quantum
systems with non-hermitean Hamiltonians enjoy a particularly
simple relation (\ref{newprop}). As a consequence, it is possible
to formulate completeness and orthonormality relations which
invoke the elements of one basis only. These relations are
inherited from the dual pair of bases providing them thus with a
sound mathematical footing. Structurally similar relations can be
derived for any pseudo-hermitean Hamiltonian.

It is a different question whether this mathematical
structure---call it `complex extension' of quantum mechanics
\cite{bender+02}, for example---is realized in nature. To draw a
positive conclusion, one would need to find a natural
interpretation of the linear, idempotent `charge operator' ${\cal
C}$. This appears difficult in the framework of non-relativistic
quantum mechanics: in spite of having real eigenvalues $s_n$ only,
the operator $\cal C$ is neither self-adjoint nor unitary while
the familiar operator of charge conjugation $\hat C$ used in field
theory is unitary.

\end{document}